\documentclass[twocolumn,showpacs,amsmath,amssymb,floatfix,prb]{revtex4}

\usepackage{graphicx}
\begin{document}

\title{Energy Transmission across Acoustically Mismatched
Solid Junctions}

\author{Jian Wang}
\author{Jian-Sheng Wang}

\affiliation{Department of Computational Science,
National University of Singapore, Singapore 117543, Republic of Singapore}

\date{3 September 2005}

\begin{abstract}
We derive expression for energy flux in terms of lattice normal
mode coordinates. Energy transmission across solid junctions from
lattice dynamic point of view is given and its relation with
atomic masses, lattice constants, and group velocities is
clarified. A scattering boundary method (SBM) is proposed for
calculating the amplitude transmission across solid junctions. The
phonon transmission coefficients and thermal conductance are
calculated for two kinds of acoustically mismatched junctions:
different chirality nanotubes $(11,0)$ to $(8,0)$, and Si-Ge
interface structure.  Our calculation shows a mode-dependent
transmission in nanotube junction due to the high symmetry
vibrating motions for nanotube atoms, indicating its possible
important role in nanotube mixture thermal conductance. Energy
transmission and Kapitza conductance across the Si-Ge interface
$[001]$ are calculated for the Si-Ge diamond-type structure. It is
shown that the energy transmission across the Si-Ge interface
depends on the incident angle and on the interface mode
conversion.  A critical incident angle about $ 42^\circ$ is
numerically found for waves incident from Ge to Si.  Our numerical
result of the Kapitza conductance at temperature $T=200\;$K is
$G_{K}=4.6\times10^{8}\; {\rm WK}^{-1}{\rm m}^{-2}$. We find
numerically scaling law $G_K \propto T^{2.87}$ for
 $[001]$ interface at low temperature.
\end{abstract}


\maketitle

\section{Introduction}
Rapid progress in the synthesis and processing of materials with
structures of nanometer length scales has created a demand for the
understanding of thermal transport in nano-scale low dimensional
devices \cite{Baowen,Jswang,K.Schwab,dgcahill}.  Recent experimental
and theoretical studies have revealed novel features of phonon
transport in these systems, such as the size-dependent anomalous heat
conduction in one-dimensional (1D) chain \cite{Jswang} and the
universal quantum thermal conductance \cite{K.Schwab}.  Thermal
transport in nanostructures may differ from the predictions of
Fourier's law based on bulk materials; this may happen because of the
existence of many acoustically mismatched interfaces in nanostructures
and because the phonon mean free path is comparable to the size of the
structure \cite{dgcahill}.  An understanding of the thermal conduction
across acoustically mismatched solid interfaces is a necessary
requirement for thermal transport engineering.

The study of thermal transport across interfaces dates back as early
as to 1940s when Kapitza resistance \cite{ETswartz} was reported and
much work has been done in this field \cite{dgcahill,ETswartz}.  In
general, theoretical modeling of this problem has been undertaken
either by the acoustic-mismatch model (AM) with scalar elastic waves,
or by the diffuse-mismatch model (DM) with Boltzmann transport
equation \cite{G-Chen}. Some numerical methods such as molecular
dynamic simulation \cite{MD} have also been used.  While AM and DM
models provide some useful reference calculations, scalar wave model
and Boltzmann transport equation are only phenomenological
descriptions and they have ignored the complexity of the
interface. Atomic-level lattice dynamic (LD) approach should be the
right way of capturing the mechanism of heat transport. Authors in
Ref.~\onlinecite{dayang} have proposed to calculate the amplitude
transmission by connecting the neighboring atoms through the dynamic
equations across the interface.  Ref.~\onlinecite{mingo} has suggested
calculating the phonon transport via Green's functions.  However, the
Green's function method gives only a formal solution for the
scattering problem. A practical calculation procedure is not provided.
Ref.~\onlinecite{dayang} used phonon mode of wave incident from one
lead to calculate the energy transmission through the amplitudes of
the transmitted/reflected waves by $t_{{\bf
k}j}={\rho_{A}\sum_{q}v_{qz}A_{q}^{2}}/{(\rho_{B}v_{0z}A_{0}^{2})}$
without any justification. How to calculate the energy transmission
from the amplitudes of the transmitted/reflected waves, and what is
its relationship with atom mass, lattice constant and group velocity?
It's not clear.  On the other hand, the well-known Landauer formula
for electronic conduction \cite{landauer} can be generalized to
thermal transport for such calculations.  Landauer formula for
ballistic heat transport across junctions has been used
\cite{gcrego,guo-hong-student-thesis} for the prediction of universal
quantum heat conductance at very low temperatures.  The energy
transmission formula derived under continuum assumption may not be
applied straightforwardly to systems on nanoscale where atomic details
are important.

In this paper, we report our results of the energy transmission
across solid junctions taking into account atomic details
from lattice dynamic point of view both theoretical and
numerically.  In Section II, we will first derive the energy flux
by lattice dynamics method and then give the energy transmission
across the interface involving atomic mass, lattice
constant and group velocity, starting from the energy conservation
law.  An analytic result for 1D acoustic mismatched chain model
from our formula is also illustrated.  We further propose
computing the amplitude transmission coefficients by solving a set
of dynamical equations with a scattering boundary condition method
(SBM) to cope with the complex interface. In Section III, we first
compute the transmission coefficients of a Carbon-nanotube
junction showing a mode-dependent transmission in nanotube
junction due to the high symmetry of vibration in nanotubes. Then
for the Si and Ge diamond-typed structure by linearized empirical
Tersoff potential, we calculate the energy transmission across the
Si-Ge interface $[001]$ to show its dependence on the incident
angle. A critical incident angle about $42^\circ$ is numerically
found for waves incident from Ge to Si.  We also calculate the
Kapitza conductance for Si-Ge interface at different temperatures
and find its value $G_{K}=4.6\times10^{8}\; {\rm WK}^{-1}{\rm
m}^{-2}$ when $T=200\,$K and numerically get its scaling law
$T^{2.87}$ at low temperature.

\section{Theory}
\subsection{Energy Flux}

In this section we derive the energy flux in terms of the normal
mode coordinates of the atomic vibration from the energy continuity
equation.  We consider systems with perfect leads on the left and
right with an arbitrary interaction at the junction.  The Hamiltonian
for each lead system (as well as the whole system) of vibrating atoms
under linear approximation takes the form
\begin{equation}
  \label{hamilton}
H = \sum\limits_l \Bigl( \sum\limits_{i,\alpha} \frac{{p_{l,i}^\alpha} ^2}{2m_i} +
\!\!\!\!\sum\limits_{l'\!,i,\alpha ;j,\beta }\!\! \frac{1}{2}
K_{l,i;l'\!\!,j}^{\alpha,\beta} u_{l,i}^\alpha u_{l'\!\!,j}^\beta  \Bigr)
 = \sum\limits_l \varepsilon_l,
\end{equation}
where $l$ or $l'$ denotes a unit cell, $i,j$ the position in a cell,
$\alpha, \beta$ the direction of vibrating motions of atoms, and
$u_{l,i}^\alpha$ the displacement from equilibrium of the atom $(l,i)$
with equilibrium position ${\bf R}_{l,i}$. The dynamic matrix
eigenvalue problem can be written as
\begin{equation}
 \label{dymatrix}
 {\bf D} \tilde {\bf e} = \omega^{2} \tilde {\bf e},\quad
 \| {\bf D} - \omega^{2} {\bf I}   \| = 0,
\end{equation}
where $ {\bf D}_{i,j}^{\alpha,\beta} = \frac{1}{\sqrt{m_i m_j}}
\sum_{l'}K_{l,i;l'\!\!,j}^{\alpha,\beta} e^{ i {\bf q} \cdot ({\bf
R}_{l'}- {\bf R}_{l} ) } $.  The eigenvector for Eq.~(\ref{dymatrix})
is $\tilde{e}_{i,n}^\alpha({\bf q)}$, where $i$ denotes the position
in a cell, $\alpha$ the direction of vibrating motion, $ n $ the
branch of polarization.  These eigenvectors satisfy
$\sum_{i,\alpha}\tilde{e}_{i,n}^\alpha({\bf q})
\tilde{e}_{i,n'}^{*\alpha}({\bf q}) =\delta_{n,n'}$.  So the vibrating
motions of atoms can be expanded into the summation of these
eigenvectors
\begin{equation}
\label{normalmode} u_{l,i}^\alpha =
\frac{1}{\sqrt{Nm_{i}}}\sum\limits_{{\bf q}, n}Q^{n}_{\bf
q}(t)\tilde{e}_{i,n}^{\alpha}({\bf q})e^{i{\bf q}\cdot {\bf R}_l},
\end{equation}
where $Q^{n}_{\bf q}(t)$ is a normal mode coordinate of the
vibration, and $N$ is the number of the unit cells. Each normal
mode represents a single harmonic oscillation $ \ddot{Q}^{n}_{\bf
q}(t) + \omega_{n}^{2}({\bf q}) Q^{n}_{\bf q}(t)=0 $. A local
energy density can be defined through the energy in cell $l$ from
Eq.~(\ref{hamilton}), as $\rho({\bf r}) = \sum_{l} \varepsilon_l
\delta({\bf r} - {\bf R}_l)$ where ${\bf R}_l$ is a lattice
vector.

Next, an expression for the energy current in the $z$ direction can be
derived from the energy continuity equation, $\partial \rho/\partial t
+ \nabla\! \cdot {\bf j} = 0$, where $ \bf j $ is the energy current
density. We transform the energy continuity equation into the momentum
space,
\begin{equation}
  \label{energycontq}
 \left\langle  \dot{\Omega}({\bf q}, t) \right\rangle_{t} +
i q_{z} \left\langle J_{z}({\bf q}, t) \right\rangle_{t} = 0,
\end{equation}
by the following Fourier transform
\begin{eqnarray}
 \label{energyfourier}
 \rho( {\bf r}, t) & = & \frac{1}{(2 \pi)^{3}} \int \Omega({\bf q},
  t)e^{i{\bf q} \cdot {\bf r} } d^{3}{\bf q}, \\
  {\bf j }({\bf r}, t) & = & \frac{1}{(2 \pi)^{3}} \int {\bf J}({\bf q},
  t)e^{i{\bf q} \cdot {\bf r} } d^{3}{\bf q}.
\end{eqnarray}
In these equations, $\Omega({\bf q},t)$ and ${\bf J}({\bf q},t)$ are
the energy density and energy current density in momentum space,
respectively, and the bracket $\left\langle \ \right\rangle_t$ denotes
the time average. Here we have assumed that the energy current
propagates along $z$ direction and so the time averaged energy current
components along other two directions equal to zero. A relation of the
total energy current in real space and in momentum space is
\begin{equation}
  \label{qlimit}
  \lim_{{\bf q} \to 0}{\bf J}({\bf q},
  t) = \int {\bf j}({\bf r}, t) d^{3}{\bf r}.
\end{equation}
Using Eq.~(\ref{energycontq}) and Eq.~(\ref{qlimit}), we can get the
average energy flux along the $z$ direction,
\begin{equation}
  \label{energyflow}
  \bar {I}_z = \frac{\left\langle J_{z}({\bf q}, t) \right\rangle_{t}}{V}
  = \frac{1}{V} \lim_{q_z \to 0}
\frac{ \left\langle {\dot {\Omega } ({\bf q} ,t)} \right\rangle _t
}{ - iq_z },
\end{equation}
where $V$ is the volume of the lead part.

With the help of Eq.~(\ref{hamilton}), Eq.~(\ref{normalmode}), and
Eq.~(\ref{energyfourier}), the energy density in momentum space is
given by
\begin{eqnarray}
 \label{energydensitynormal}
\Omega \left({\bf q}_0 ,t\right) =
\sum_l \varepsilon_l e^{ - i {\bf q}_0 \cdot {\bf R}_l }
=\frac{1}{2}\!\!\!\!\sum_{{\bf q}, {\bf q}', n, n'}\!\!\! \Big\{ \big(
\dot{Q}_{{\bf q}}^n \dot{Q}_{{\bf q}'}^{n'} \nonumber\\
  +\;\omega_{n'}^{2}({\bf q}')
{Q}_{{\bf q}}^n {Q}_{{\bf q}'}^{n'} \big)
 \times \sum_{i,\alpha} \tilde
{e}_{i,n}^\alpha({\bf q}) \tilde {e}_{i,n'}^{*\alpha}({\bf q}')
\delta_{{\bf q} + {\bf q}', {\bf q}_0}\Big\},
\end{eqnarray}
where $Q_{\bf q}^n$'s are normal mode coordinates for the $n$-th
branch phonons.  Combining Eq.~(\ref{energyflow}) and
Eq.~(\ref{energydensitynormal}), the time-averaged energy flux along
$z$-direction is
\begin{equation}
  \label{energyflowq}
  \bar {I}_z =\frac{-i}{V}\sum\limits_{{\bf q},n}
{\omega_{n}({\bf q}) \frac{\partial \omega_{n}({\bf q})}{\partial q_z} \langle
Q_{\bf q}^{n} \dot{Q}_{\bf q}^{n*}  \rangle_t}.
\end{equation}
Note that during the derivation of Eq.~(\ref{energyflowq}), ${\Omega }
({\bf q} ,t)$ should be first differentiated with respect to time and
then takes the limit $q_{z}\rightarrow 0$.  Eq.~({\ref{energyflowq}})
is the formula for the energy flux in terms of normal mode coordinates
from lattice dynamic point of view. It provides the basic starting
point for studying the energy transmission across the interface.

\subsection{Energy Transmission}
Once the energy flux formula Eq.~({\ref{energyflowq}}) is obtained, we
can further discuss the problem of energy transmission across the
interface.  The idea for lattice dynamic approach to study energy
transmission across the interface was first proposed in
Ref.~\onlinecite{dayang}. Assuming an eigen mode lattice wave for
Eq.~(\ref{dymatrix}) incident from one lead, it will be refracted by
the interface. The reflected and transmitted waves are assumed the
linear combination of the eigen mode lattice waves of
Eq.~(\ref{dymatrix}) at each side of the lead.  These amplitude
transmissions are solved through the equations of motion for the
boundary atoms.  But it should be noted that these amplitude
transmissions are not energy transmissions.  What are their relation
to the energy transmissions?  As already cited in the introduction
part of this paper, several assumptions \cite{dayang} are made to
calculate the energy transmissions from these amplitude transmissions.
Apart from this problem, any eigenvector for Eq.~(\ref{dymatrix}) can
be multiplied by a whatever coefficient and it is still the
eigenvector.  It means that the choice of eigenvector is not unique
because of the linear nature of the harmonic system.  This
indeterminacy will complicate the calculation of the amplitude
transmissions and makes the amplitude transmission coefficients
non-unique. However the energy transmissions are determined by
the system and have unique values.  The key to this problem is energy
conservation, which means that the energy flux from the left and right
lead should be equal. In this part, we use the energy flux
Eq.~({\ref{energyflowq}}) derived from the energy continuity equation
in the previous section to clarify these issues.

We assume that the vibration of the atoms for each
incident/tranmitted normal mode is
\begin{equation}
\label{eigenmode} \tilde {u}_{l,i,n}^\alpha (\omega, {\bf q} ) =
\frac{1}{\sqrt {m_i } }\tilde {e}_{i,n}^\alpha ( {\bf q})
e^{i({\bf q} \cdot {\bf R}_l - \omega t)},
\end{equation}
where $\tilde {e}_{i,n}^\alpha ( {\bf q})$ is the eigenvector for
dynamic matrix Eq.~(\ref{dymatrix}) and it satisfies
$\sum_{i,\alpha}\tilde{e}_{i,n}^\alpha({\bf q})
\tilde{e}_{i,n'}^{*\alpha}({\bf q}) =\delta_{n,n'}$.  It is
obvious that our choice for the eigen mode is unique and
deterministic. Compared with the choice of incident normal mode in
Ref.~\onlinecite{dayang}, the difference is that the  atomic mass
should be included.  The reason for this is simple: the
eigenvectors of Eq.~(\ref{dymatrix}) are not the travelling wave
solution for the system and its relation to travelling wave is to
be divide by $\sqrt{m_i}$ according to the definition of
Eq.~(\ref{dymatrix}).    We assume the same form,
Eq.~(\ref{eigenmode}), for all the incident, reflected, and
transmitted waves.  When one particular mode $\tilde
{u}_{l,i,n}^{\alpha,L} (\omega, {\bf q} ) $ is incident from the
left lead, it will be scattered by the interface. So the reflected
wave is given as $\sum_{n'}t^{LL}_{n'n} \tilde
{u}_{l,i,n'}^{\alpha,L} (\omega, {\bf -q'} ) $ and the transmitted
wave is $\sum_{n'}t^{RL}_{n'n} \tilde {u}_{l,i,n'}^{\alpha,R}
(\omega, {\bf q''} ) $. In these equations,
$t_{n'\,n}^{\sigma'\,\sigma}$ is the amplitude
transmission/reflection coefficients from mode $n$ in the lead $
\sigma=L $ to mode $n'$ in lead $\sigma'=R, L$.  Note that $\bf
q'$ and ${\bf q}''$ satisfies $\omega=\omega_{n'}({\bf -q'}) =
\omega_{n'}({\bf q}'')$. Similar expression can be written down
for the right lead. The total wave for one particular lead, say
the left lead, is an arbitrary superposition of all these eigen
modes. Thus, the motion of the atoms is described by the wave:
\begin{eqnarray}
  \label{wave}
  \Psi _{l,i}^{\alpha ,L} \left( {\omega ,t} \right) &=& \sum_n
a_n^L \tilde {u}_{l,i,n}^{\alpha ,L}(\omega, {\bf q})  \nonumber \\
&+&\!\! \sum_n \Bigl( \!\!\sum_{n',\sigma = L,R} \!\!
a_{n'}^\sigma t_{n\,n'}^{L\,\sigma}  \Bigr) \tilde
{u}_{l,i,n}^{\alpha ,L}(\omega , -{\bf q}'),
\end{eqnarray}
where $a_n^\sigma$ is an arbitrary amplitude of the mode $n$ in lead
$\sigma$.  With the help of our newly derived energy flux
Eq.~(\ref{energyflowq}) for lattice dynamics, after transforming
Eq.~(\ref{wave}) into normal mode coordinates, we can get the heat
flux for the lead $\sigma = L, R$:
\begin{eqnarray}
  \label{energyflowqclassical}
  \bar {I}^\sigma  =
\frac{1}{V_\sigma}\sum\limits_n {\left( | {a_n^\sigma } |^2
  - \Bigl| {\sum\limits_{n',\sigma '}
{a_{n'}^{\sigma '} t_{nn'}^{\sigma \sigma '} } } \Bigr|^2
\right ) \omega_n^{2} \frac{\partial \omega_{n}^\sigma({\bf q})}{\partial
q_z}}.
\end{eqnarray}
Since the energy must be conserved, there is no net energy
accumulation in the junction, which means that the time-averaged
energy currents from both sides are equal, ${I}^L \equiv {I}^R$, where
the energy currents ${I}^{\sigma}= \bar{I}^{\sigma}\cdot S_{\sigma}$,
$ \sigma=L,R$ and $ S_{\sigma}$ represents the cross area for each
lead. This condition leads to the following identity for the
transmission amplitudes,
\begin{equation}
  \label{transmission}
  \sum\limits_{\sigma ,n}{ t_{nn^{\prime}}^{\sigma\sigma^{\prime}}
  t_{nn^{\prime\prime}}^{*\sigma\sigma^{\prime\prime}}\tilde
  {v}_n^\sigma}
  = \tilde {v}_{n^{\prime}}^{\sigma^{\prime} } \delta _{n^{\prime}\sigma
^{\prime},n^{\prime\prime}\sigma ^{\prime\prime}},
\quad \tilde {v}_n^\sigma = {v_n^\sigma }/{\emph{l}_{z}^\sigma },
\end{equation}
where $\emph{l}_{z}^\sigma$ is the length of unit cell along the $z$
direction, $v_n^\sigma = \partial \omega^\sigma_n/\partial q_z$.  This
equation is analogous to the unitarity condition for the scattering
matrix in electronic transmission.  To clearly understand the meaning
of Eq.~({\ref{transmission}}), we can write it into a matrix product
equation ${\bf t}^{\dag} {\bf \tilde{v }}{\bf t}= {\bf \tilde{v}}$
with
\begin{subequations}
\label{vtdeifnition}
\begin{eqnarray}
{\bf \tilde{v}} & =  & \left(
\begin{array}
[c]{cccccc}%
\tilde{v }_1^L & 0 & 0 & 0 & 0 & 0\\
0 & \tilde{v }_2^L & 0 & 0 & 0 & 0 \\
0 & 0 & \cdots & 0 & 0  & 0\\
0 & 0 & 0 & \tilde{v }_1^R & 0  & 0\\
0 & 0 & 0 & 0 & \tilde{v }_2^R  & 0\\
0 & 0 & 0& 0 & 0 & \cdots
\end{array} \right), \\
{\bf t} &=& \left(
\begin{array}
[c]{cccccc}%
t^{LL}_{11} & t^{LL}_{12} & \cdots & t^{LR}_{11} & \cdots   &  \\
t^{LL}_{21} & t^{LL}_{22} & \cdots & t^{LR}_{21} &\cdots    &  \\
\cdots  & \cdots  & \cdots & \cdots & \cdots   &  \\
t^{RL}_{11} & t^{RL}_{12} & \cdots & t^{RR}_{11} & \cdots   &  \\
t^{RL}_{21} & t^{RL}_{22} & \cdots & t^{RR}_{21} & \cdots   &  \\
\cdots  & \cdots  & \cdots & \cdots & \cdots   &
\end{array} \right).
\end{eqnarray}
\end{subequations}
The above expression and the technique we used are similar to the
continuum case,\cite{guo-hong-student-thesis} but the mathematics
involved in the lattice dynamic framework is simpler and
clearer. Several interesting conclusions can be made from
Eq.~(\ref{transmission}) and Eq.~(\ref{vtdeifnition}).

\textit{(1)} The amplitude transmissions from each lead are not
independent.  They must obey Eq.~(\ref{transmission}) so that the
energy will be kept conserved. It means that which lead is chosen for
calculating the energy transmission does not matter. They will give
the same energy flux from either of the lead. In practice, this should
be a good criterion to check the validity of the algorithm in
calculating the amplitude transmission.

\textit{(2)} From the diagonal terms of Eq.~(\ref{transmission}), the
energy transmission $\tilde{T}_{n'n}^{\sigma'\sigma}$ from mode
$(\sigma , n) $, which means $\sigma$ lead and $n$ branch, to the mode
$ (\sigma', n')$ is given by
\begin{equation}
  \label{energytran}
 \tilde{T}_{n'n}^{\sigma'\sigma}= |t_{n'n}^{\sigma'\sigma}|^{2}\frac{\tilde{v}^{\sigma'}_{n'}}{\tilde{v}^{\sigma}_{n}}.
\end{equation}
The total reflected and transmitted energy transmission for $(\sigma ,
n) $ is given by
\begin{subequations}
\begin{eqnarray}
  \label{reftran}
 \mathcal{R}_{n}^{\sigma}= \sum _{n'}|t_{n'n}^{\sigma
 \sigma}|^{2}\frac{\tilde{v}^{\sigma}_{n'}}{\tilde{v}^{\sigma}_{n}},
 &&
 \mathcal{T}_{n}^{\sigma}= \!\!\sum _{n', \sigma' \neq
 \sigma}\!\!\!|t_{n'n}^{\sigma'
 \sigma}|^{2}\frac{\tilde{v}^{\sigma'}_{n'}}{\tilde{v}^{\sigma}_{n}}, \\
\mathcal{R}_{n}^{\sigma} + \mathcal{T}_{n}^{\sigma}  \equiv  1.&&
\end{eqnarray}
\end{subequations}

\textit{(3)} The energy transmission's relation to atomic mass,
lattice constant and the group velocity is explicitly expressed in
Eq.~(\ref{reftran}).  It can be seen that the atom mass $m_{i}$
for each lattice site ${\bf R}_{l,i}$ should be considered in each
eigen mode wave as in Eq.~(\ref{eigenmode}).
Ref.~\onlinecite{dayang} used $t_{{\bf
k}j}={\rho_{A}\sum_{q}v_{qz}A_{q}^{2}}/(\rho_{B}v_{0z}A_{0}^{2})$
to calculate the energy transmission from the amplitude
transmission $A_{q}$. This equation is correct only if the atomic
masses are the same in a unit cell.  In actuality, the ratio of
Silicon mass density to Germanium mass density is about $ 0.4375 $
at temperature $T=300\;$K, while the accurate coefficient should
be $ 28/72\approx0.389$ if the energy is conserved and lattice
constants are same for both leads. Ref.~\onlinecite{zhao} improved
this by using the atomic mass ratio in the final energy
transmission to keep energy conserved. But when the unit cell of
system is composite,  the choice of incident mode wave of
Ref.~\onlinecite{dayang} is wrong.  It should be chosen as our
Eq.~(\ref{eigenmode}).  So the results of thermal conductance
between metals and $BaF_2$ in Ref.~\onlinecite{dayang} are
arguable using the method in Ref.~\onlinecite{dayang}. The lattice
constant is also considered in Eq.~(\ref{reftran}), which may have
significant contribution when the difference in the lattice
constants for the left and right lead crystal is large.

\textit{(4)} Comparing with the continuum case \cite{gcrego}, we see
an extra factor of the lattice dimension in $z$ direction,
$\emph{l}_{z}^\sigma$.  This factor can be included in the amplitude
transmission if we redefine our normal mode as that in
Eq.~(\ref{eigenmode}) multiplied by $\sqrt{V_{\rm cell}}$, where
$V_{\rm cell}$ is the volume of a unit cell.

Next, we quantize Eq.~(\ref{energyflowq}) with
\begin{subequations}
\label{quantization}
\begin{eqnarray}
\hat {Q}_{\bf q}^n (t)\!\!\! & = \!\!\!&  \sqrt {\frac{\hbar
}{2\omega _n}} \left( \hat {a}_{n,{\bf q}} e^{-i\omega _nt} + \hat
{a}^\dag _{n, -
{\bf q}} e^{  i\omega_nt} \right), \\
\!\!\! \dot {\hat {Q}}_{-{\bf q}}^n (t)\!\!\! & =\!\!\! & -i\sqrt
{\frac{\hbar \omega _n}{2}} \left( { \hat {a}_{n, -{\bf q}} e^{-
i\omega _n t} - \hat {a}^\dag _{n,{\bf q}} e^{i\omega _n t}}
\right).
\end{eqnarray}
\end{subequations}
We get the quantization of the heat current $ \bar{I} = \frac{1}{V}
\sum_{{\bf q},n} \hbar \omega_n({\bf q}) v_n({\bf q}) \hat{a}^\dag_{n,
{\bf q}} \hat{a}_{n,{\bf q}} $. Similarly after quantizing
Eq.~(\ref{energyflowqclassical}) with the help of the relation
Eq.~(\ref{transmission}), and taking thermal average, we can get the
Kapitza conductance as
\begin{eqnarray}
\label{current}
 G_{K} &=&  \lim_{\Delta T \to 0} \frac{\bar{I}}{\Delta T } \nonumber \\
  &=&\!\! \frac{1}{V}\!\!\!\! \sum \limits_{{\bf q}, n, v_n>0}\!\!\!\! \hbar
 \omega_{n}({\bf q})v_{n}({\bf q}) \mathcal{T}_{n}({\bf q},
 \omega_{n}) \frac{\partial f(\omega_n, T)}{\partial T},
\end{eqnarray}
where $\mathcal{T}_{n}({\bf q}, \omega_{n}) =
\sum_{n'}|t_{n'n}^{RL}|^{2}\tilde{v}^{R}_{n'}/\tilde{v}^{L}_{n} $ and
$f(\omega_n, T)$ is the Bose-Einstein distribution at temperature $T$.
The dispersion relation and group velocity refer to the left
lead.  In deriving Eq.~(\ref{current}), we have used the fact that
phonons in the lead obey the Bose-Einstein distribution.

\subsection{Scattering Boundary Method}

\begin{figure}[bt]
\includegraphics[width=0.9\columnwidth]{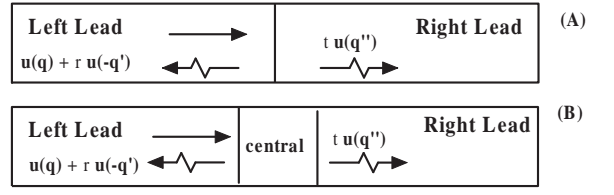}
\caption{\label{fig:schematic} Schematic show for scattering
boundary condition.}
\end{figure}

The central issue now is to have an efficient method to calculate the
amplitude transmission coefficients across junctions at the atomistic
level.  Ref.~\onlinecite{dayang} has proposed solving the amplitude
transmissions through connecting boundary atoms for abrupt solid
interface.  But this appears to be a difficult problem for general
complex interfaces. As illustrated in Fig.~\ref{fig:schematic}, the
central part of atoms is not known so the method proposed in
Ref.~\onlinecite{dayang} will not work.  Transfer matrix method can be
used for simple 1D models with a few atoms in the central part to
obtain analytical solutions.  But this method is not numerically
stable for a long 1D system (such as a disordered 1D system) or 3D
systems because the errors get amplified by continuous multiplications
of a matrix with eigenvalues larger than 1.  In fact, we have tried
transfer matrix method for the nanotube junction and found large
numerical errors of 30\% or more.

Here we propose what we call the scattering boundary equation method
(SBM) as a solution.  Each atom in the system satisfies the dynamic
equation, $-m_i \omega ^2 {\bf u}_i + \sum\nolimits_j K_{ij} ( {\bf
u}_i - {\bf u}_j ) = 0 $, where the $3\times 3$ matrix $K_{ij}$ is the
force constants between atom $i$ and $j$.  The boundary conditions are
of the form $ {\bf \tilde u }_{q,n'} + \sum_n r_n { \bf \tilde
u}_{-q,n}$ for the incoming and reflected waves and $\sum_{n} t_n {
\bf \tilde u'}_{q',n}$ for the outgoing waves, where ${ \bf \tilde
u}_{q,n}$ and ${\bf \tilde u'}_{q',n}$ are the eigen modes on the left
and right leads, while the reflection and transmission coefficients
$r_n$ and $t_n$ are unknown. These equations in matrix form are
illustrated as
\begin{equation}
  \label{sbe}
\left( {{\begin{array}{*{10}c}
 \hfill & \cdots \hfill  & \hfill & \hfill \\
 \vdots \hfill & { K_{ii} } \hfill & { -
K_{ij} } \hfill & \hfill & \hfill \\
 \hfill  & \hfill & \hfill & \hfill \\
 1 \hfill & &  & { - \tilde u_{ - q,n} } \hfill & \hfill \\
 \hfill & \hfill & 1 \hfill & \hfill & { - \tilde u'_{ q',n} } \hfill \\
\end{array} }} \right)\left( {{\begin{array}{*{10}c}
 { u_{1\!\!\!\!\!} } \hfill \\
 \vdots \hfill \\
 {u_{m\!\!\!\!\!} } \hfill \\
 r_{n} \hfill \\
 t_{n} \hfill \\
\end{array} }} \right) = \left( {{\begin{array}{*{10}c}
 0 \hfill \\
 \vdots \hfill \\
 0 \hfill \\
 { \tilde u_{q,n\!\!\!\!\!} } \hfill \\
 0 \hfill \\
\end{array} }} \right),
\end{equation}
where $K_{ii}=-m_i \omega^{2}+\sum\nolimits_j {K_{ij} }$.  These
equations can be both analytically and numerically solved by the
conventional method if the matrix is square.  In some cases for higher
dimensions, however, the boundary conditions are complicated and the
number of equations may be larger than that of variables. But these
equations are not linearly independent.  Nevertheless, these equations
are consistent and simultaneous.  Under this condition the scattering
boundary equations can still be solved numerically by the standard
singular value decomposition method \cite{numericalrecipes}.

\subsection{Analytic Results on 1D Model}
To demonstrate the formulas obtained in the previous part and the
scattering boundary equation method, we give some analytic results on
a toy 1D model.  More realistic material solid junctions, e.g.
nanotube conjunction and Si-Ge interface, will be discussed in the
next section.
\begin{figure}[bt]
\includegraphics[width=1.0\columnwidth]{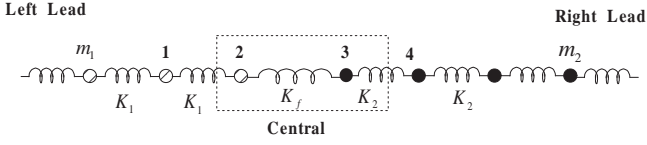}
\caption{\label{fig:schematic2} Schematic show for 1D acoustically
mismatched chain model.}
\end{figure}
We consider two acoustically mismatched chains of different atomic
masses $m_1, m_2$, connected with springs of different stiffness $K_1,
K_2$, and lattice constant $a, b$, on each side as illustrated in
Fig.~\ref{fig:schematic2}. In the central part they are connected by a
spring with stiffness $K_f$. The dynamic matrix for the 1D chain gives
the eigenvector $\tilde{e}_{l}=e^{iqx_{l}}$ and
$\omega^{2}_L=\frac{2K}{m}\bigl(1-\cos(qa)\bigr)$.  From
Eq.~(\ref{eigenmode}), the incident, reflected and transmitted waves
from the left lead can be written as
\begin{subequations}
\label{leftlead}
\begin{eqnarray}
  u_{Lr} &=& \left(\frac{1}{\sqrt{m_{1}}}e^{iq_{1}x}+t^{LL}\frac{1}{\sqrt{m_{1}}}e^{-iq_{1}x}
  \right) e^{-i\omega t},
  \\
 u_{Lt} &=& t^{RL} \frac{1}{\sqrt{m_{2}}} e^{iq_{2}x}e^{-i\omega t},
\end{eqnarray}
\end{subequations}
where, following the notation of Eq.~(\ref{wave}), $ t^{LL} $ and $
t^{RL} $ are the reflected and transmitted amplitudes.  Similar
expressions can be written down for the wave incident from the right
lead as
\begin{subequations}
\label{rightlead}
\begin{eqnarray}
  u_{Rr} &=& \left(\frac{1}{\sqrt{m_{2}}}e^{-iq_{2}x}+t^{RR}\frac{1}{\sqrt{m_{2}}}e^{iq_{2}x}  \right)e^{-i\omega t},
  \\
 u_{Rt} &=& t^{LR} \frac{1}{\sqrt{m_{1}}} e^{-iq_{1}x}e^{-i\omega t}.
\end{eqnarray}
\end{subequations}
As illustrated in the schematic Fig.~\ref{fig:schematic2}, there
are two atoms, labelled $2$ and $ 3 $ in the figure,  in the
connecting part. The equations of motion for the two atoms are
\begin{subequations}
\label{eqmotion}
\begin{eqnarray}
  (-m_{1}\omega^{2}+K_{1}+K_{f})u_{2} - K_{1}u_{1}-K_{f}u_{3} &=&
  0,  \\
 (-m_{2}\omega^{2}+K_{f}+K_{2})u_{3} - K_{f}u_{2}-K_{2}u_{4} & = &
  0.
\end{eqnarray}
\end{subequations}
For wave incident from the left lead, the SBM equations can be
expressed as
\begin{equation}
  \label{leftsbm}
\left( \begin{array} [c]{llllll} 
\!\! -K_1& K_{22} \! &  -K_{f} & \! 0& \!  0&0  \\
0& \! \!\!\!\!\! - K_{f}\! \! & K_{33}\! \! &\! \!\! \!-K_{2}& 0  & 0 \\
1& 0  & \! 0 &\! 0 &\! \!\! \! \!\! \! \frac{-e^{-iq_{1}x_{1}}}{\sqrt{m_1}} & 0 \\
0& 1 &\!  0 &\! 0 & \! \!\!\! \!\! \! \frac{-e^{-iq_{1}x_{2}}}{\sqrt{m_1}} & 0 \\
0 & 0  & 1 & 0 &\!  0 & \! \!\!\!\!\!\!\!\! \!\! \!\! \frac{-e^{iq_{2}x_{3}}}{\sqrt{m_2}} \\
 0 & 0 & 0 & 1 &\!   0 & \! \!\!\!\!\!\! \!\! \! \! \!\! \frac{-e^{iq_{2}x_{4}}}{\sqrt{m_2}}
\end{array} \right ) \!\! \!\!
\left( \begin{array} [c]{c}
u_1  \\
u_2 \\
u_3 \\
u_4  \\
t^{LL} \\
t^{RL}
\end{array} \right )\!\!=\!\!
\left( \begin{array} [c]{c}
0  \\
0 \\
\frac{e^{iq_{1}x_{1}}}{\sqrt{m_1}} \\
\frac{e^{iq_{1}x_{2}}}{\sqrt{m_1}} \\
0\\
0
\end{array} \right ),
\end{equation}
where $K_{22}=-m_{1}\omega^{2}+K_{1}+K_{f}$,
$K_{33}=-m_{2}\omega^{2}+K_{f}+K_{2}$ and $q_{1}, q_{2}$ satisfy that
$\omega_L(q_{1})=\omega_R(q_2)$. For simplicity, we only consider the
condition that the two spring chains are connected directly by the
spring $K_{f}$. If there are more atoms at the central part, we only
need to add equations like Eq.~(\ref{eqmotion}) for these atoms to
Eq.~(\ref{leftsbm}). But the scattering boundary condition in
Eq.~(\ref{leftsbm}), last four rows, does not change.  From
Eq.~(\ref{leftsbm}), we can get
\begin{subequations}
\label{leftrt}
\begin{eqnarray}
& t^{LL}& =
  \bigl[-K_fe^{iq_1x_2}+( \frac{K_{33}}{K_{f}} -  \frac{K_{2}}{K_{f}}e^{iq_2b} )  \nonumber \\
& &(K_{22}e^{iq_1x_2} - K_1e^{iq_1x_1} )\bigr] / \bigl[(\frac{K_{33}}{K_f}-
\frac{K_2}{K_f}e^{iq_2b}) \nonumber \\
& &(K_{22}e^{-iq_1x_2}-K_1e^{-iq_1x_1})
+ K_f e^{-iq_1x_2} \bigr],\\
&t^{RL}& = \sqrt{\frac{m_2}{m_1}}\bigl[(K_{22}e^{iq_1x_2}-K_1e^{iq_1x_1})\nonumber \\
& & +(K_{22}e^{iq_1x_2}-K_1e^{iq_1x_1})\bigr]/K_f.
\end{eqnarray}
\end{subequations}
Similarly for wave incident from the right lead, we have SBM
equations
\begin{equation}
  \label{rightsbm}
\left( \begin{array} [c]{llllll} 
\!\! -K_1& K_{22} \! &  -K_{f} &  0& \!\! 0  \, & \, \, 0  \\
0& \! \!\!\!\!\! - K_{f}\! \! & K_{33}\! \! &\!  \! \!-K_{2}& \!\! 0 \, & \, \, 0 \\
1& 0  & \! 0 &\!\! \!\! 0 & \! \!  0 & \! \!\!\!  \frac{-e^{-iq_{1}x_{1}}}{\sqrt{m_1}}  \\
0& 1 &\!  0 &\! \! \!\!0 &\! \!   0 &  \! \!\!\!   \frac{-e^{-iq_{1}x_{2}}}{\sqrt{m_1}}  \\
0 & 0  & 1 &\! \!\! 0 &\!   \!\!\!\!\!\! \!  \frac{-e^{iq_{2}x_{3}}}{\sqrt{m_2}}&0  \\
 0 & 0 & 0 & \! \!\!1  &  \! \!\! \! \! \!\!
 \frac{-e^{iq_{2}x_{4}}}{\sqrt{m_2}}  & 0
\end{array}\!\! \! \right ) \!\! \!\!
\left( \begin{array} [c]{c}
u_1  \\
u_2 \\
u_3 \\
u_4  \\
t^{RR} \\
t^{LR}
\end{array} \right )\!\!=\!\!
\left( \!\! \!\! \begin{array} [c]{c}
0  \\
0 \\
0\\
0\\
\frac{e^{-iq_{2}x_{3}}}{\sqrt{m_2}} \\
\frac{e^{-iq_{2}x_{4}}}{\sqrt{m_2}}
\end{array} \!\! \!\! \right ).
\end{equation}
From Eq.~(\ref{rightsbm}), we can also get $t^{RR}$ and $t^{LR}$.
Next we construct Eq.~(\ref{vtdeifnition}) for 1D model as
\begin{eqnarray}
  \label{onedvt}
{\bf t} =\left( \begin{array} [c]{cccc}
 t^{LL} & t^{LR} \\
 t^{RL} & t^{RR}
  \end{array} \right ),  \,\,\,\,\,
  {\bf v} =\left( \begin{array} [c]{cccc}
 \frac{v^{L}}{a} & 0 \\
 0 & \frac{v^{R}}{b}
  \end{array} \right ).
\end{eqnarray}
Matrix Eq.~(\ref{vtdeifnition}), ${\bf t}^{\dag} {\bf \tilde{v }}{\bf t}=
{\bf \tilde{v}}$, gives four equations for this 1D model,
\begin{subequations}
\label{onedfoure}
\begin{eqnarray}
|t^{LL}|^{2}\frac{v^{L}}{a}+|t^{RL}|^{2}\frac{v^{R}}{b} =
\frac{v^{L}}{a}, \\
|t^{RR}|^{2}\frac{v^{R}}{b}+|t^{LR}|^{2}\frac{v^{L}}{a} =
\frac{v^{R}}{b}, \\
t^{LL*}t^{LR}\frac{v^{L}}{a} + t^{RL*}t^{RR}\frac{v^{R}}{b}=0,  \\
t^{LL}t^{LR*}\frac{v^{L}}{a} + t^{RL}t^{RR*}\frac{v^{R}}{b}=0.
\end{eqnarray}
\end{subequations}
These equations can be easily verified using the results from
Eqs.~(\ref{leftsbm}, \ref{leftrt}, \ref{rightsbm}). These four
equations also guarantee the same energy transmission both for the
left and the right lead, that is
\begin{equation}
\label{trand1}
\mathcal{T}^{L}=\mathcal{T}^{R}=
|t^{RL}|^{2}\frac{\frac{v^{R}}{b}}{\frac{v^{L}}{a}}=
|t^{LR}|^{2}\frac{\frac{v^{L}}{a}}{\frac{v^{R}}{b}}.
\end{equation}
It can be easily seen from the simple 1D case that amplitude
transmissions from the left and right lead are not independent and
should satisfy the relation~(\ref{transmission}) to keep the energy
conserved.

Furthermore, for 1D quantum thermal energy current, with the help of
the relation~(\ref{transmission}), after quantizing
Eq.~(\ref{energyflowqclassical}), we get the formula
\begin{equation}
  \label{landuerequation}
\bar {I} = \frac{1}{2\pi }\sum\limits_n {\int_{\omega _{\min }
}^{\omega _{\max } }\!\!\!\!\! d \omega\;{\hbar \omega \Bigl(f
(\omega ,T_L ) - f (\omega ,T_R )\Bigr)T_n (\omega )} },
\end{equation}
where $T_{n}\left( \omega \right) =\sum\limits_{m}{\left\vert
{t_{mn}^{RL}}\right\vert ^{2}{\tilde{v}_{n}^{R}}/{\tilde{v}_{m}^{L}}}$
is the energy transmission probability, $f(\omega, T_\sigma)$ is the
Bose-Einstein distribution for the left or right lead.
Eq.~(\ref{landuerequation}) is the Landauer formula for quantum
thermal energy flow in one dimension.

\section{Simulation}
In this section, we report our results for two kinds of solid
junctions: nanotube junction and the Si-Ge interface. Nanotube has
been found to be good candidate for thermal conduction materials and
the thermal conduction has been improved through adding nanotube
mixture\cite{nanotube}.  But the interface between nanotubes are
inevitable. How they will affect thermal conduction becomes an
interesting problem.  Thermal conduction across the Si-Ge interface
has been studied by Ref.~\onlinecite{dayang} using the simplified $
fcc $ lattice model and recently by Ref.~\onlinecite{zhao} in diamond
structure with equal lattice constants for Si and Ge and with $ fcc$
neighboring atoms considered. In our paper, we will consider more
realistic crystal structure for Si and Ge to investigate the energy
transmission dependence on incident angle and low-temperature scaling
behavior of Kapitza conductance.

\subsection{Nanotube Junction}

\begin{figure}[bt]
\includegraphics[width=0.9\columnwidth]{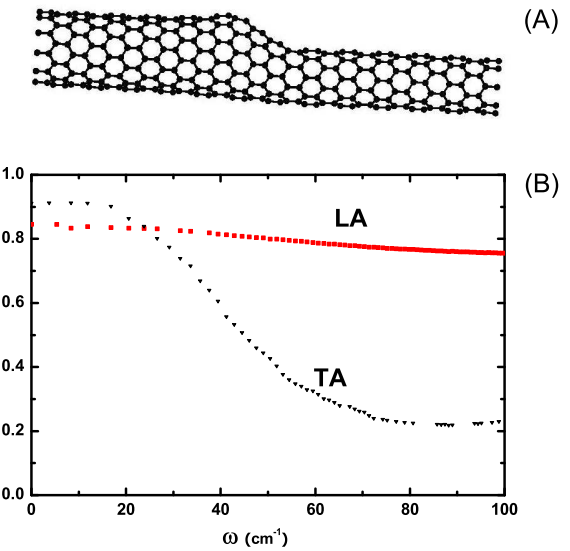}
\caption{\label{fig:nanotube1}(a) Structure of a (11,0) and (8,0)
nanotube junction. (b) The energy transition $T_n(\omega)$ as a
function of angular frequency $\omega$.}
\end{figure}

\begin{figure}[bt]
\includegraphics[width=0.9\columnwidth]{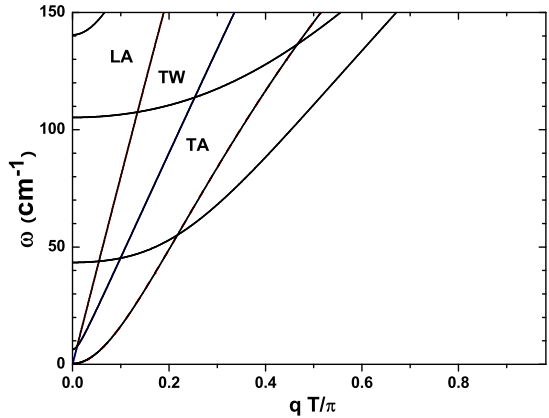}
\caption{\label{fig:nanotube2} Phonon dispersion for nanotube
$\mbox{(11,0)}$. The wave number $q$ is expressed in terms
of the length of nanotube translation vector $T$.}
\end{figure}

The semiconductor nanotube junction structure (11,0) and (8,0) is
first constructed by a geometrical method as in
Ref.~\onlinecite{nanotubebook}. However there are many defects in the
geometrically constructed structure. Next we optimize the junction
structure by a second-generation Brenner potential \cite{dwbrenner} to
let the atoms get their equilibrium positions.  The optimized
structure is shown in Fig.~\ref{fig:nanotube1}(a). Then under small
displacement, the force constants are derived numerically from the
same potential.  The phonon dispersion calculated from these
linearized force constants for nanotube $\mbox{(11,0)}$ is illustrated
in Fig.~\ref{fig:nanotube2}, in which four acoustic branches are
considered for energy transport: the longitudinal mode (LA), doubly
degenerate transverse mode (TA), and the unique twist mode (TW) in
nanotubes.

Following the SBM method, the SBM equations are constructed and it
involves $ 504 $ equations and $455$ variables.  However, some
equations are not linearly independent and are consistent because the
motion for each atom is connected with the motions of others forming
travelling waves and they are confined together to construct the
nanotube structure.  We also numerically verified the consistency of
these equations through the calculation of ranks for SBM
equations. The rank we get is equal to $455$ within numerical
accuracy, which is equal to the number of variables. These equations
are solved through the standard singular value decomposition method
\cite{numericalrecipes}.

We then discuss the SBM results of the transmission coefficients for a
nanotube junction, shown in Fig.~\ref{fig:nanotube1}(b).  Although all
modes of a given frequency are considered, we did not find mode-mixing
or mode-conversion behavior among acoustic modes at the lower
frequency range. For example, the reflected and the transmitted waves
across the junction for the incident LA mode waves are both only LA
modes. We think that this is due to the high symmetrical properties of
atomic motion for nanotubes.  Each travelling wave on nanotube has its
own symmetrical property resulting from the symmetry of the nanotube
structure.  The symmetry on the left and right lead is different.  It
is hard to convert one specific high symmetrical motion to the other
different symmetrical motion through the conjunction except when the
symmetrical properties on each lead are the same.  We think that this
kind of mismatch in the symmetry of motion for nanotubes will play an
important role in the thermal conduction of nanotube mixtures.

The LA modes are common symmetrical motion for both the left and right
lead.  The transmission for LA mode stays around 0.8 with only small
changes.  This value is below the AM model prediction \cite{walittle}
of $0.98$ with the longitudinal group velocity $20.18\; {\rm km}\!
\cdot\! {\rm s}^{-1}$ and $20.95\; {\rm km}\!  \cdot\! {\rm s}^{-1}$
for $\mbox{(11,0)}$ and $\mbox{(8,0)}$. In contrast, the transmissions
of the TW mode and many other optical modes are nearly zero or very
small. This appears related to the difference in rotational symmetries
of these modes.  The transmission for TA mode decreases with
frequency.  This can be accounted for two reasons: (1) Although the TA
mode is a common symmetry for both left and right lead, the kind of
transverse symmetry is destroyed by the central junction part. It can
be seen from Fig.~\ref{fig:nanotube1} that the central junction part
does not have transverse symmetry.  (2) There is a mismatch of
dispersion relation for TA mode because the TA mode has a nearly
quadratic dispersion relation, as illustrated in
Fig.~\ref{fig:nanotube2}. These reasons will affect the transmission
for TA mode at high frequency when the vibration motion of atoms is
fast.

So it can be seen from above that energy transmission across nanotube
junctions is strongly dependent on symmetry property of the lead
nanotube. This symmetry dependence is also observed by molecular
dynamics.\cite{nanotubemd} Usually such mismatch in symmetry will
reduce the number of possible transmission modes and thus blocks the
thermal conduction across the junction.  We also propose that this
kind of mode-dependent transmission behavior may be important for
further application such as phonon filters.

\subsection{Si-Ge Interface}
The Kapitza conductance across the Si-Ge interface have been calculated by
lattice dynamic method in Ref.~\onlinecite{dayang} using the
simplified $ \mbox{\textit{fcc}} $ lattice model and recently by
Ref.~\onlinecite{zhao} in diamond structure with $ \mbox{\textit{fcc}}
$ neighboring atoms considered. However, the relation of energy
transmission to the incident wave angle and the temperature dependence
of Kapitza conductance at lower temperature remains unclear. Here we
use our newly derived formulas combined with the simulation to get an
understanding of these problems in more realistic Si-Ge structure. We
first discuss the energy transmission's relation to the incident
angles and to wave mode conversion at the interface.  The Kapitza
conductance across Si and Ge interface is also calculated and its
scaling law in relation to temperature at low-temperature regime is
numerically analyzed.

\begin{figure}[b]
\includegraphics[width=0.9\columnwidth]{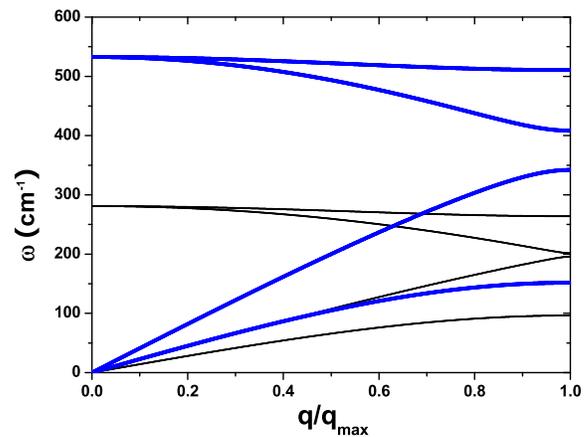}
\caption{\label{fig:dispersion} Phonon dispersion along $\Gamma-L$ for
Si and Ge. The thick lines are for Si and the thin lines for Ge. The
wave number is in terms of the maximum value along $\Gamma-L$
direction.  }
\end{figure}

The lattice structure for Si and Ge crystal we considered is the
diamond structure. The atomic mass for Si and Ge are $28 \,
\mbox{\textit{amu}}$ and $ 72.61 \, \mbox{\textit{amu}}$.  The lattice
constants for each conventional unit cell are $a=5.43\,$\AA\ and
$a=5.658\,$\AA, respectively.  We use the Tersoff potential
\cite{dwbrenner} under small displacement to get the linearized force
constants.  By Tersoff potential truncation function,\cite{dwbrenner}
only the nearest 4 atoms are considered for each atom with the
distance $ \sqrt{3}a/4 $.  We choose the primary cell consisting of
two atoms $(0,0,0)$ and $(a/4, a/4, a/4)$ to construct the dynamic
matrix. The phonon dispersion for Si and Ge along the $\Gamma-L$
direction can be calculated through the linearized force constants
from the dynamic matrix and are illustrated in
Fig.~\ref{fig:dispersion}.  The maximum LA branch along $\Gamma-L$
frequency for Si and Ge are $343\, {\rm cm}^{-1}$ and $ 196\, {\rm
cm}^{-1} $. It can be seen that the linearized force constants for the
nearest atoms well reproduce the Si and Ge phonon
dispersion.\cite{giannozzi} The group velocity is calculated through
\begin{equation}
\label{groupvelocity} v_g = \frac{1}{2\, \omega} \frac{{\bf
\tilde{e}}^{\dag} \frac{\partial {\bf D} }{\partial q}{\bf \tilde{e}},
}{{\bf \tilde{e}}^{\dag}\cdot {\bf \tilde{e}} },
\end{equation}
where $\bf D$ is the dynamic matrix and ${\bf \tilde{e}}$ the
eigenvector as shown in Eq.~(\ref{dymatrix}).  The group velocity of
LA branch phonons along the $\Gamma-L$ direction for Si and Ge is
about $7.8\,{\rm Km}/{\rm s}$ and $4.4\,{\rm Km}/{\rm s}$.

For the Si or Ge diamond structure, there are six modes of vibration
as illustrated in Fig.~\ref{fig:dispersion} , among which three
branches are acoustic and other three ones are optical.  Using
Eq.~(\ref{eigenmode}) when one mode of wave $ \tilde
{u}_{l,i,n}^\alpha (\omega, {\bf q} ) = \frac{1}{\sqrt {m_{{\rm Si}} }
}\tilde {e}_{i,n}^\alpha ( {\bf q}) e^{i({\bf q} \cdot {\bf R}_l -
\omega t)} $ is incident from the lead, it has three modes of
reflected wave and three modes of transmitted wave. Due to linearity
of the system, the frequency does not change during refraction. We
choose the Si-Ge interface with normal in $[001]$ direction. Since the
system is homogeneous in the $x$ and $y$ direction, the reflected
waves and the transmitted waves have the same momentum in these
directions $q'_{x}, q''_{x}=q_{x}$; $q'_{y}, q''_{y}=q_{y}$. $q'_{z}$
and $q''_{z}$ is found to satisfy $\omega({\bf q})=\omega_{{\rm
Si}}({\bf q'})=\omega_{{\rm Ge}}({\bf q''})$. If $q'_{z}$ or $q''_{z}$
is complex, ${\rm Im}[q'_{z}]$ or ${\rm Im}[q''_{z}]$ should be
negative.  The force constants for the Si-Ge interface are chosen as $
K_{ij}=(K_{ij}^{Si}+K_{ij}^{Ge})/2 $. Each atom in the incident lead
part has the motion ${\bf u}_{\rm reflect}= {\bf \tilde{u}}({\bf q}) +
r_{1}{\bf \tilde{u}}({\bf -q'_{1}}) + r_{2}{\bf \tilde{u}}({\bf
-q'_{2}})+r_{3}{\bf \tilde{u}}({\bf -q'_{3}})$; similarly for the
motion in the transmitted wave ${\bf u}_{\rm transmit}= t_{1}{\bf
\tilde{u}}({\bf q''_{1}}) + t_{2}{\bf \tilde{u}}({\bf
q''_{2}})+t_{3}{\bf \tilde{u}}({\bf q''_{3}})$. Connecting the Si-Ge
interface provides us two equations of motion,
\begin{eqnarray}
\label{motion}
 \bigl(-m_{Si}\omega^{2}+\sum \limits_{j}{\bf K}_{ij}\bigr){\bf u}_{i}-\sum \limits_{j}{\bf K}_{ij}{\bf
 u}_{j} & =  & 0, \\
 \bigl(-m_{Ge}\omega^{2}+\sum \limits_{j}{\bf K}_{ij}\bigr){\bf u}_{i}-\sum \limits_{j}{\bf K}_{ij}{\bf
 u}_{j} & = & 0.
\end{eqnarray}
There are all $8$ different atoms in the above two equations. Each
atom should have the form of solution of the reflected ${\bf u}_{\rm
reflect}$ or the transmitted ${\bf u}_{\rm transmit}$, which depends
on its position. The scattering boundary equations are
constructed. There are $30$ equations (two equations of motion give
$6$ and another $8$ atoms taking the scattering boundary condition
solution give $24$) and $30$ variables ($8$ atoms give $24$ and $6$ for
reflected coefficients ${\bf r}$ and transmitted
coefficients ${\bf t}$) in the scattering boundary equations. This set
of equations can be solved by a conventional method. When ${\bf r}$ and
${\bf t}$ is obtained, the energy reflection coefficients $R$ and
transmission coefficients $T$ for each mode are given by summation of
Eq.~(\ref{energytran}).
\begin{figure}[b]
\includegraphics[width=0.95\columnwidth]{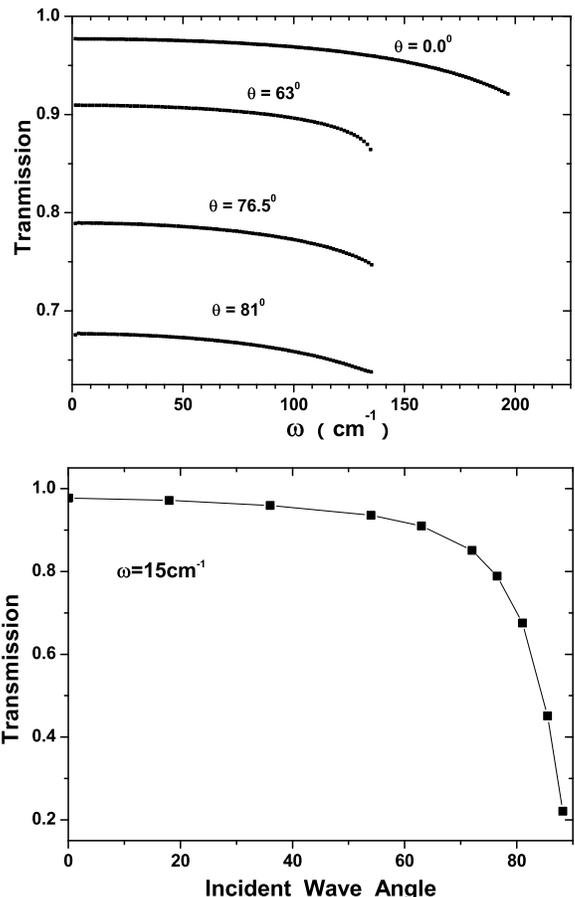}
\caption{\label{fig:angle1} Dependence of energy transmission on the
incident angle. LA phonon incident from Si to Ge. }
\end{figure}

\begin{figure}[b]
\includegraphics[width=0.95\columnwidth]{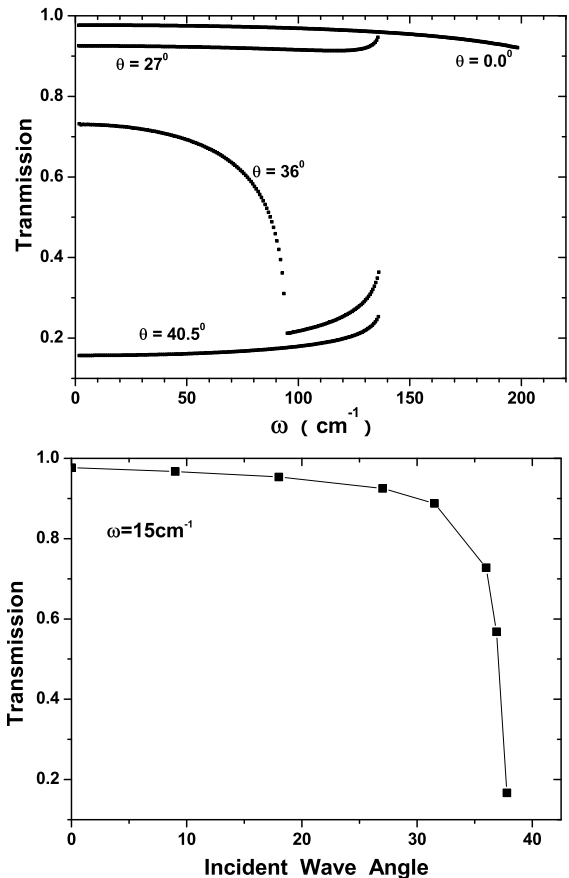}
\caption{\label{fig:angle2} Dependence of energy transmission on the
incident angle. LA phonon incident from Ge to Si.  }
\end{figure}

We first report the result of the dependence of energy transmission on
the incident angle.  A continuum wave incident on the surface whose
wave vector makes an angle $\theta$ with the surface normal is
refracted in accordance with Snell's law.  Does the refraction of
phonon incident on the solid interface observe the Snell's law? Is
there any new character for phonon refraction?  We calculated the
energy transmission for $LA$ mode incident from Si to Ge and incident
from Ge to Si. The results are illustrated in Fig.~\ref{fig:angle1}
and Fig.~\ref{fig:angle2}.

It can be seen from Fig.~\ref{fig:angle1} that when the angle $\theta$
for the phonon incident from Si to Ge increases, energy transmission
decreases slowly. But once the angle is larger than $80^{\circ}$, the
transmission decreases rapidly. In contrast to the transmission from
Si to Ge, $LA$ mode wave transmission from Ge to Si shows a rich
character as illustrated in Fig.~\ref{fig:angle2}. For wave with the
incident angle $\theta \neq 0 $, the transmission first decreases with
the increase of frequency. But when the frequency goes over a certain
value, for example $\omega=93\, {\rm cm}^{-1}$ for
$\theta=36^{\circ}$, the transmission begins to increase. This
behavior can be understood by the mode conversion at the interface. It
can be seen from Fig.~\ref{fig:dispersion} that the maximum frequency
for $TA$ modes in Ge is about $98\, {\rm cm}^{-1}$.  When the
frequency is below this value, there are $TA$ modes for the reflected
wave; but over this value, the reflected wave can not be converted
into $TA$ modes. So due to lack of reflected modes, the transmission
increases.

Furthermore, for $LA$ waves incident from Ge to Si, there exists a
critical value, about $42^{\circ}$, above which transmission equals
$0$. We can estimate the critical angle with the help of Snell's law
for continuum wave. The group velocities for normal incident waves in
direction $[001]$ are $v_{g}^{Si}\approx 6.87\,$Km/s and $v_{g}^{Ge}
\approx 3.78\,$Km/s.  The Snell's law gives the critical angle from Ge
to Si $ \theta_{c}=\sin^{-1}(v_{g}^{Ge}/v_{g}^{Si})=
33.4^{\circ}$. This value is a little lower than the observed critical
angle value.  Nevertheless, we think that the Snell's law gives a rough
estimation of the critical value for the critical incident angle for
phonons.

When the incident waves from Si to Ge and from Ge to Si are both
normal to the surface ($\theta = 0 $), the energy transmission
from both sides are about the same $0.98$. We use the acoustic
mismatch model \cite{walittle} to estimate
$4Z_{Si}Z_{Ge}/(Z_{Si}+Z_{Ge})^{2}\approx0.97$.  It can be seen
that the acoustic mismatch model describes the transmission for
normal incident waves well.

\begin{figure}[b]
\includegraphics[width=0.9\columnwidth]{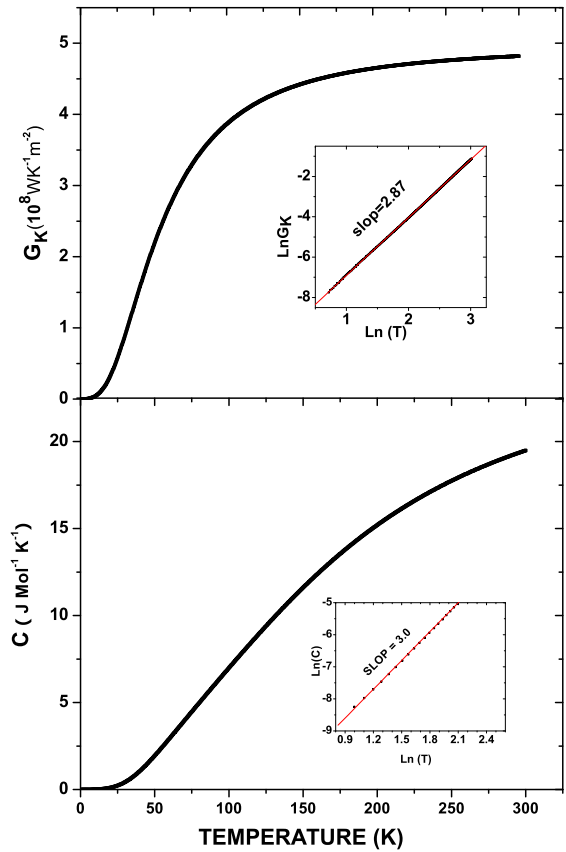}
\caption{\label{fig:conductance} Kapitza conductance and specific heat
from linearized Tersoff potential. }
\end{figure}

The Kapitza conductance with change of temperature is calculated using
Eq.~(\ref{current}). The results for Kapitza conductance and heat
capacity using our linearized force constants are illustrated in
Fig.~\ref{fig:conductance}.  The Kapitza conductance for Si-Ge $[100]$
interface calculated from our model is $G_{K}=4.6\times10^{8}\, {\rm
WK}^{-1}{\rm m}^{-2}$ when $T=200\,$K. When the temperature goes over
$200\,$K, we find that the Kapitza conductance changes little with the
temperature and is saturated.  For comparison, we plotted the heat
capacity in Fig.~\ref{fig:conductance}.  It can be seen from the
figure that the heat capacity continues to increase with the
temperature when $T>200\,$K.  In comparison with the heat capacity,
the saturation of Kapitza conductance can be accounted by little
contribution of energy transmission from high frequencies.
At low temperatures, as illustrated in inset of
Fig.~\ref{fig:conductance}, the Kapitza conductance scales as
$T^{2.87}$, while the heat capacity scales as $T^{3}$ in
accordance with the Debye model.  We have sampled enough points in
the first Brillouin zone to ensure that Kapitza conductance and
heat capacity converge numerically.  However, due to the small
deviation from 3, we cannot rule out the possibility that the
exponent for Kapitza conduction is also 3.
Ref.~\onlinecite{dayang} reported that Kapitza conductance scales
as $T^{3}$ at low temperatures for $fcc$ interface irrespective of
the properties for the left and right lead. The Kapitza
conductance's dependence on temperature is an intriguing problem
\cite{ETswartz}, with much experimental work done in this field.
Most experiments gave $T^{\alpha}$ with $\alpha\leq3$ for solid
interface as reviewed in Ref.~\onlinecite{ETswartz}. So far no
experimental result is available for the temperature dependence of
the Si-Ge interface. What value should it be is an interesting
problem.  Compared with the results of Ref.~\onlinecite{dayang},
our discrepancy from $T^{3}$ comes from the anisotropy of the
energy transmission because of the diamond structure is used for
the calculation of the transmission,  while
Ref.~\onlinecite{dayang} took the isotropic assumption in their
calculation.

\section{Conclusion}
In summary, we derive expressions for energy flux in terms of
lattice normal mode coordinates. Energy transmission across solid
junctions from lattice dynamic point of view is given and its
relation with atom masses, lattice constants, and group velocities
is also clarified. A scattering boundary method (SBM) is proposed
for calculating the amplitude transmission across solid junction.
Our calculation shows a mode-dependent transmission in nanotube
junction due to the high symmetry vibrating motion for nanotube
atoms, indicating its possible important role in nanotube mixture
thermal conductance. Energy transmission and Kapitza conductance
across the Si-Ge interface are calculated for the Si-Ge
diamond-type structure from linearized Tersoff potential. It is
shown that the energy transmission across the Si-Ge interface
depends on the incident angle and on the interface mode
conversion.  A critical incident angle about $ 42^{\circ} $ is
numerically found for waves incident from Ge to Si.  The critical
angle in the reverse direction is much larger. The Kapitza
conductance saturates to about $G_{K}=4.6\times10^{8}\, {\rm
WK}^{-1}{\rm m}^{-2}$ for $T>200\,$K. We numerically get its
scaling law $T^{2.87}$ for $[001]$ interface at low temperature.

In additions, we remark that

(1) Eq.~(\ref{transmission}) and Eq.~(\ref{vtdeifnition}) are the
basic relations for amplitude transmissions in linear systems,
irrespective of the details of junction part. The results for any
algorithm to calculate amplitude transmission from lattice point of
view should satisfy these relations to keep the energy conserved. It
dose not matter which lead to be chosen for the energy transmission
because the energy flux will be the same according to
Eq.~(\ref{transmission}).  So Eq.~(\ref{transmission}) provides a
criterion for checking the validity of the algorithm.

(2) Lattice dynamic approach to solid junction is only under linear
approximation, which may be accurate at low temperatures.  However,
for high temperature the nonlinear effect or anharmonic effect has to
be considered.  How to include the effect of nonlinearity into phonon
scattering at the solid junction will be an interesting problem.

\section{Acknowledgements}
This work is supported in part by a Faculty Research Grant of National
University of Singapore. We thank Lin Yi and Nan Zeng for discussions.

\end{document}